\documentclass[aps, pre, twocolumn, showpacs, superscriptaddress, floatfix]{revtex4}

\usepackage{graphics} 
\usepackage{graphicx}
\usepackage{epsfig}
\usepackage{amsmath,amssymb,wasysym,epsfig,capt-of,ifthen,calc}
\usepackage{bbm} 
\usepackage{latexsym} 
\usepackage{setspace} 
\usepackage{array} 
\usepackage{delarray} 
\usepackage{afterpage} \usepackage{graphicx}
\usepackage{dcolumn}
\usepackage{bm}
\newcommand{\be}{\begin{equation}}
\newcommand{\ee}{\end{equation}}
\newcommand{\bea}{\begin{eqnarray}}
\newcommand{\eea}{\end{eqnarray}}

\begin{document}

\title {Random Sequential Renormalization of Networks: Application to Critical Trees}
\author{Golnoosh Bizhani} 
\affiliation{Complexity Science Group, University of Calgary, Calgary, Canada}
\author{Vishal Sood} \affiliation{Niels Bohr Institute, Copenhagen, Denmark}
\author{Maya Paczuski} 
\affiliation{Complexity Science Group, University of Calgary, Calgary, Canada}
\author{Peter Grassberger}
\affiliation{Complexity Science Group, University of Calgary, Calgary, Canada}
\affiliation{NIC, Forschungszentrum J\"ulich, D-52425 J\"ulich, Germany}

\date{\today}

\begin{abstract} 
We introduce the concept of random sequential renormalization (RSR) for arbitrary networks. 
RSR is a graph renormalization procedure that locally aggregates nodes to produce a 
coarse grained network.  It is analogous to the (quasi-)parallel renormalization schemes 
introduced by C. Song {\it et al.} [C. Song {\it et.al.,} Nature (London) {\bf 433}, 392 (2005)] and studied 
by F. Radicchi {\it et al.} [F. Radicchi {\it et al.,} Phys. Rev. Lett. {\bf 101}, 148701 (2008)], but much simpler 
and easier to implement.  Here we apply RSR to critical trees and derive analytical
results consistent with numerical simulations. Critical trees exhibit three regimes in their 
evolution under RSR. (i) For $N_0^{\nu}\lesssim N<N_0$, where $N$ is the number 
of nodes at some step in the renormalization and $N_0$ is the initial  size of the tree,  RSR is described by a mean-field theory, and fluctuations from one realization to another 
are small. The exponent $\nu=1/2$ is derived using random walk and other arguments. The degree distribution 
becomes broader under successive steps, reaching a power law $p_k\sim 1/k^{\gamma}$ 
with $\gamma=2$ and a variance that diverges as $N_0^{1/2}$  at the end of this regime. Both 
of these latter results  are obtained from a scaling theory. (ii) For 
$N_0^{\nu_{\rm star}}\lesssim N \lesssim N_0^{1/2}$, with $\nu_{\rm star}\approx 1/4$ 
hubs develop, and fluctuations between 
different realizations of the RSR are large. Trees are short and fat with an average radius that is ${\cal O}(1)$. Crossover functions exhibiting finite-size 
scaling in the critical region $N\sim N_0^{1/2} \to \infty$ connect the behaviors in the 
first two regimes. (iii)  For $N \lesssim N_0^{\nu_{\rm star}}$, 
star configurations appear with a central hub surrounded by many leaves.  The distribution of
stars is broadly distributed over this range.
The scaling behaviors found under RSR are identified with a continuous transition  in a  process called ``agglomerative percolation'' (AP), with the coarse-grained nodes in RSR corresponding to clusters in AP that grow by simultaneously attaching to all their neighboring clusters.
\end{abstract}
\maketitle

\section{Introduction}

Renormalization  is a basic concept in statistical physics.  It is a process whereby 
degrees of freedom in a system are successively eliminated by coarse graining.  At the same time 
system parameters are rescaled to compensate for the decimation, and the smallest scale is reset 
to its original value~\cite{amit}. Since a series of such transformations is itself a transformation, 
the transformations $\{{\cal R}\}$ form a semi-group: the  ``renormalization group'' (RG).  

If the system is statistically invariant  under $\{{\cal R}\}$, one speaks of RG invariance. An invariant system exhibits an asymptotic  fixed point under the RG flow with
scaling described by homogeneous functions.  Prototypical RG fixed points are critical phenomena 
displayed at continuous phase transitions as for the Ising model, by a-thermal systems like  directed~\cite{Hinrich} or ordinary~\cite{Stauffer} 
percolation, relativistic quantum field theories~\cite{Zinn}, or the 
Feigenbaum (period doubling) cascade in one-dimensional dynamical systems~\cite{Feigenbaum}. Systems with 
the same fixed point under RG are in the same universality class and share the same critical 
exponents.

It is natural to ask if similar concepts can be applied to glean meaningful information
about complex networks. A positive answer was suggested in Ref.~\cite{Song} and has stirred much 
interest. In the present paper we start an investigation to further explore whether and in what sense 
this can be true.

For models on a lattice, coarse graining can be accomplished either in Fourier space or 
in real space.  A typical real space RG proceeds heuristically by covering a 
spin lattice with a regular grid of boxes, and replacing the degrees of freedom in each box 
by a ``super-spin"~\cite{Stauffer}. Interactions between spins in 
neighboring boxes are used  to specify the couplings between super-spins.

However, many real world phenomena are better represented as complex networks rather than regular 
lattices. Although research in this area has exploded in recent years (for reviews see, 
e.g., Refs.~\cite{newmanRev, BA, boccaletti}), our understanding of the statistical physics of complex 
networks has not caught up with the vast body of knowledge accrued over decades for lattice systems.
Some phase transitions on networks (e.g., in the spreading of epidemics~\cite{newman,pastor})
are straightforward generalizations of critical phenomena on lattices. 
Yet it is not clear whether the RG, and real-space renormalization, in particular,
can be applied systematically to complex networks.

Closely related to renormalization is the notion of  fractal dimensions~\cite{amit,Zinn}. 
Many complex networks are {\it small world} networks~\cite{Milgram, Watts}, where the number 
of nodes within reach of any node via paths of length $r$ increases exponentially with $r$. 
Via any standard definition, this gives infinite fractal dimensions. However Song 
{\it et al.}~\cite{Song}, made claims to the contrary, finding finite fractal dimensions 
for several real-world networks based on a quasi-parallel renormalization scheme. 
A real-space RG for networks that is {\it not based} on the concept of fractal 
dimensions,  but studied in terms of the flow under renormalization, was proposed by Radicchi 
{\it et al.}~\cite{Radi1,Radi2}. 

A fundamental issue pertinent to all the work up to now on renormalization of networks 
(see, for instance, Refs.~\cite{Song,Radi1,Radi2,Song2,Song3,Kim1,Kim2,Kim3}) is that completely 
covering a network with equal size boxes leads to a number of  unavoidable dilemmas that
could lead to erroneous conclusions. Conceptually, covering the system with boxes of equal 
sizes is a flagrant violation of the original idea of Hausdorff~\cite{Falconer}, where 
the system ought to be covered with a partitioning whose elements have individually 
optimized sizes up to some largest size $r$. In most applications this is not a serious 
impediment, and a covering with equal size elements gives equivalent results. Thus most 
estimates of fractal dimensions in physics use fixed box sizes, although there are well 
known cases where this leads to erroneous results. The most famous one is given by any 
infinite but countable set of points, which according to Hausdorff, but not according 
to any covering algorithm with fixed box size, has zero dimension.

One reason why this problem can be neglected in many physical systems is that the number
of points per box (or, more precisely, the weight of each box) has small fluctuations,
in particular, relative to a distribution whose width increases exponentially with 
box size. For small world networks, where, indeed, the {\it maximum} number of nodes 
increases exponentially, the schemes of Refs.~\cite{Song,Song2,Song3,Kim1,Kim2,Kim3} may 
give misleading results because {\it most} boxes have only a few nodes. 
Then the problems associated with fixed box size become acute  and there is no reason 
to believe that the results obtained are  related to genuine 
fractal dimensions of the underlying graph.

Even with fixed box size, the covering should also be optimized with respect to the 
exact placement or tiling of the boxes, which is an NP hard problem~\cite{Song3}. 
Heuristic methods for this optimization have been claimed to work~\cite{Song,Song2,Kim3}, 
but as a matter of fact they depend on the order in which boxes are laid down. Thus 
they are not true {\it parallel} substitutions of nodes by super-nodes, but 
quasi-parallel since the single step of tiling the whole network is implemented as 
a sequence of partial tilings. Combined with the problem of almost empty boxes, this 
means that the efficiency of the box covering algorithm changes both within each 
renormalization step (the boxes put down first contain in general more vertices than 
later boxes), and from one step to the next. 

Another problem with the (quasi)parallel renormalization scheme is that each step of 
renormalization dramatically reduces the number of nodes in the network. Therefore 
few points and less statistics are obtained for analyzing renormalization flow. 
This becomes particularly serious in the case of small world networks which collapse 
to one node in a few steps, even when the initial network size is huge. This has 
been overcome to some extent in Ref.~\cite{song10} by performing a renormalization where 
only parts of the network are coarse-grained at each step, at the cost of adding 
more parameters and making the results harder to interpret.

In view of these problems, we decided to study graph renormalization for unweighted, 
undirected networks by means of a purely {\it sequential} algorithm: At each step one
node is selected at random, and all nodes within a fixed distance of it (including itself) 
are replaced by a single super-node. The super-node has links to all other nodes that 
were connected  to the original subset absorbed into the super-node. This is repeated 
until the network collapses to a single node.

Our method avoids the problem of finding an optimum tiling as well as problems with 
almost empty boxes. A further advantage of our random sequential renormalization (RSR) 
procedure is that each step has a much smaller effect on the network, and thus the whole
renormalization flow consists of many more single steps for a finite system and allows 
for a more fine grained analysis.

If there are fixed points underlying this RG flow, then they will manifest themselves in 
terms of (finite-size) scaling laws, which hold for large initial networks at intermediate 
times.  Here time is measured by the number of steps in the RSR.  At intermediate times, 
the system is far from both the initial network and the non-invariant final network 
composed of a single super node.

On any graph, including networks or lattices, the super-nodes can be viewed as  clusters that grow by attaching to all of their neighboring clusters, up to a distance $b$ in the network of clusters.  This process, called ``agglomerative percolation'', has been solved exactly in one dimension and shown to exhibit scaling laws with exponents that depend on $b$~\cite{Son-2010}.  On a square lattice in two dimensions, critical behavior is seen which is in a different universality class~\cite{claire} than ordinary percolation.  Thus the scaling behavior seen in RSR occurs as a result of a type of percolation transition and is not restricted to cases where the underlying graph is fractal.

Here
we apply our RSR methodology to critical trees and also
find evidence for a critical point (which is, however, {\it not} a fixed point of the 
RSR!) where the number of links attached to any node 
(i.e., its degree) follows  a power law and  divergences appear for e.g. the variance of the 
degree distribution.  The size of the networks at the transition point diverges as $N_0^{1/2}$, slower than the initial network size ($N_0$) in the limit of infinite system size.  Below this transition,
renormalized trees are short and fat with an average depth (or radius) which is ${\cal O}(1)$.
We determine some critical exponents using random walk and other arguments, as well as a mean-field 
theory for the initial, uncorrelated phase. We use, in addition, the observation that all 
renormalized networks for $b=1$ eventually reach a star dominated by a central hub before they 
collapse to a single node. Our results are confirmed by means of finite-size scaling 
analyses of results from numerical simulations. These simulations also reveal scaling 
behavior for the probability distribution for the sizes of networks that first reach a 
star configuration.  This turns out to be equivalent to the distribution of sizes one step 
before the network collapses to a single node.  Stars first appear for renormalized networks 
when the size of the network  is $N\lesssim N_0^{\nu_{\rm star}}$ with $\nu_{\rm star}\approx 1/4$.  

In Sec. II, we define the general RSR procedure for any network as well
as the specific ensemble of networks we analyze in this paper. Section III presents 
our theoretical and numerical results for RSR of critical trees.  Finally,  we end 
with conclusions and outlook for future work in Sec. IV.

\section{The Model} 

\subsection{Random Sequential Renormalization}
\begin{figure}
\includegraphics[width=0.9\columnwidth]{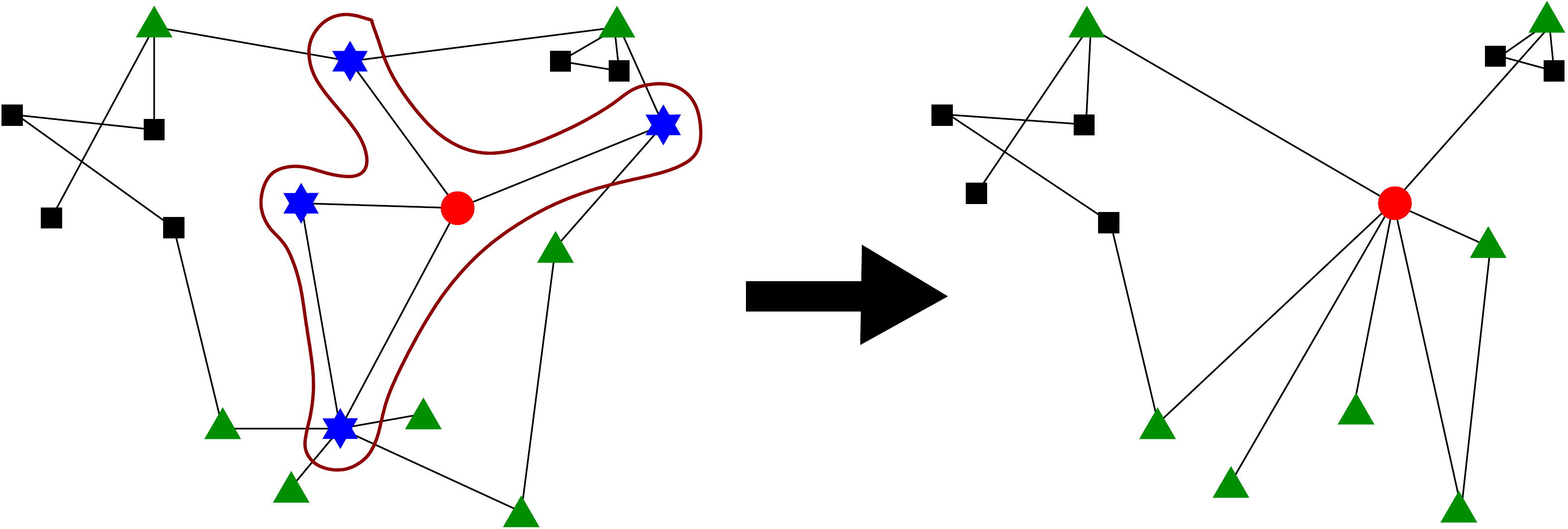}
\caption{(Color online) One step of RSR with $b=1$. The randomly chosen target node (red circle), 
absorbs all its nearest neighbors (blue stars).  All links to the absorbed nodes 
(from green triangular nodes) are then redirected to the target. Alternatively one can view the super-node as a cluster (bounded by the red curve) that subsequently grows by invading its neighboring clusters.}
\label{fig:renorm}
\end{figure}
For any undirected, unweighted  graph, RSR with {\it radius} 
$b$ ($b = 1,2, \ldots$) is defined as follows: Starting with a graph with $N_0$ nodes, we 
produce a sequence of graphs of strictly decreasing sizes $N_t$ with $0<t\leq T$ and  $N_T=1$.
For each step $t\to t+1$ ($t$ is called ``time'' in the following): \\
(i) We choose randomly a  {\it target node} $i \in [1,\ldots, N_t]$.\\
(ii) We delete all nodes that can be reached from $i$ by at least one path of length $1\leq \ell \leq b$. \\
(iii) We also delete all links {\it between} these chosen nodes, and all links connecting them to $i$.\\
(iv) Each link connecting any node outside this neighborhood to a deleted node is redirected towards
the target. 
(v) If this creates a multiple link between any two nodes, it is replaced by a single link.

Hence the target node $i$ is replaced by a super-node that maintains all links to the outside. Its 
internal features, however, are erased from the network, consistent with coarse graining.
Figure~\ref{fig:renorm} shows an example of one step 
of RSR for $b=1$. After absorbing its neighbors the super-node is treated 
like any other node and the process repeats until the network  collapses into a single node.
One could also vary the probability of choosing a target node by a function of its mass (the number of nodes absorbed into it), or its degree (the number of links attached to it), but these aspects
 are not explored here.

When $b=1$, only nearest neighbors of the target node are deleted. For $b>1$ each step can be 
implemented by performing $b$ successive decimations with radius one on the same target.
Although this method is slightly slower than an optimal coding where all nodes within distance 
$\leq b$ of the target are found and deleted in a single step, it reduces code complexity and 
potential sources of errors. 

For any radius $b\geq 1$, RSR exhibits two  trivial fixed points: a graph consisting of a single 
node, and an infinitely long chain. For a long but finite chain, the time until a single node 
is reached is $T=\lceil N_0/2b\rceil$.  In one dimension, the exact probability to find any consecutive sequence of node masses for any $N_0$ and at any time has been determined~\cite{Son-2010}. At late times, and for large $N_0$ the mass distribution of the nodes exhibits scaling both at small and large sizes with (different) exponents that depend on $b$.  For $b=1$ another fixed point exists, which is a star 
with  infinitely many leaves. In that limit, the probability to choose the central hub of the star as 
the target vanishes. With probability one, a single leaf is removed during each RSR step. 
For a finite number $N_{\rm star}-1$ of leaves, a star has an average life time $\bar T= {\cal O}(N_{\rm star})$ before it collapses into a single node.  Notice that simple stars are not 
fixed points for $b>1$, as any star reduces to a single node in one step with probability one.
In this paper we study only the case of RSR with $b=1$. 

\subsection{Initial graph ensemble} 

The ensemble of critical trees is generated as follows:  Starting with a single node,
each node can have 0, 1, or 2 offspring with 
probabilities $1/4$, $1/2$ and $1/4$.  (Hence the mean number of offspring is 1.)
The process runs until it dies due to fluctuations. The sizes of trees obtained in this way are distributed according to an  inverse power law $P(N_0) \sim N_0^{-3/2}$~\cite{Stauffer}. From these  we pick a large  ($\approx 10^2-10^3$) ensemble of trees with the desired (large) $N_0(\pm10\%)$, and discard all others.  Note that simply truncating trees that survive up to $N_0$ would give a biased sampling of the ensemble. 

 This construction  generates a rooted tree, with important consequences for joint degree distributions of  adjacent nodes.
The direction of growth leaves its imprint on them. For ordinary undirected random graphs (Erd\"os-Renyi graphs), it is well known that the degree distribution for pairs of nodes obtained by randomly
choosing a link is different from that obtained by choosing any two nodes at random. If the degree 
distribution is $p_k$, the distribution of degree pairs for linked nodes is not $p_k p_{k'}$,
but $ kk'p_k p_{k'}/\langle k\rangle^2$, because  higher degree nodes have a greater chance of being 
attached to a randomly chosen link. For the present model, two connected nodes are always in a 
mother - daughter relationship. In particular, all nodes have in-degree one; that is, they have one mother (except for the root). If $k$ is the {\it out}-degree of the mother and $k'$ the {\it out}-degree of the  daughter, then the distribution of degree pairs obtained by randomly choosing
links is 
\be
   { kp_k p_{k'} \over \sum_{l,l'}lp_lp_{l'}} = {kp_kp_{k'} \over \langle k\rangle} \quad .
\ee
While high degree mothers have a greater chance of appearing in a pair than low degree mothers,
no such bias holds for daughters. Otherwise said, if we pick a random node, the out-degrees of 
its daughters will be distributed according to $p_{k'}$, while  the out-degree of its mother
is distributed $\propto kp_k$. Notice that this implies that our ensemble of critical 
trees is {\it not} equivalent to the ensemble of critical Erd\"os-Renyi graphs.

In the following, we shall always denote by $p_k$ the distribution of out-degrees, and 
we will, for simplicity, always call $k$ the ``degree" (even though the real degree is $k+1$).
 
\section{Analytical Calculations and Simulation Results}

\subsection{Evolution of the tree size, $N$}\label{treesize}

Let $n_{k}$ be the number of nodes with degree $k$, and $N=\Sigma_k n_k$ the total number of 
nodes in the tree (i.e., its size, at a given step).
Both $N$ and $n_{k}$ are fluctuating functions of time $t$. 
Since target nodes are picked randomly, the average degree of the target is $\langle k \rangle
\equiv N^{-1}\Sigma_k kn_k = 1-1/N$, where the last equality follows from the fact that the 
total number of links in a tree is always $N-1$. Since all the target's neighbors (both its mother, unless it is the root, and any daughters) are deleted in the 
subsequent renormalization step, we get the exact result
\be
   \overline{\frac{\Delta N}{\Delta t}}= -\langle k \rangle -1 +{1\over N} = -2+{2\over N} \quad .
\label{eq:N}
\ee
Here the overline denotes an average over the randomness of the  last step only,
while brackets denote ensemble averages (except for $\langle k \rangle$) 
including also the randomness from previous RSR steps. 
Approximating $t$ by a continuous variable and performing such an ensemble average
gives \begin{equation}
   \langle N\rangle = N_0-2t+\ln\left({N_0-1\over \langle N\rangle-1}\right)\;.
   \label{N_evol}
\end{equation}
(The integration can only be performed for $N>1$.)  We have replaced  
$\langle 1/N \rangle$ on the right hand side of Eq.~(\ref{N_evol}) by $1/\langle N\rangle$, which is a mean-field approximation. We show in Sec.~\ref{random_walk_sec} that this mean-field regime extends up to a time
when $N\sim {\cal O}(N_0^{1/2})$.

\subsection{Evolution of the degree distribution}

The probability that a randomly chosen node in a network has degree $k$ is  $p_k=n_k/N$. 
The change of $ n_k$ in one step of renormalization has three contributions,
\begin{equation}
  \overline{\frac{\Delta n_k}{\Delta t}}= r_k+s_k +q_k,        \label{dndt}
\end{equation}
where:
\begin{itemize}
\item $r_k$ is a loss term associated with the possibility that the target had  (old) degree $k$ before
the considered renormalization step. 
It is  
\be
   r_k = -p_k \quad .                                               \label{r}
\ee
\item $s_k$ is a loss term from the (old) neighbors of the target having degree $k$. Assuming
no degree correlations, which is also a mean-field approximation, 
and summing over all (old) degrees $k'$ of the  target gives 
\bea
 s_k &=& -\sum_{k'}k' p_{k'}  p_k -\sum_{k'}p_{k'}  \Biggl({k p_k\over \sum_l lp_l} \Biggr)\nonumber \\
     &=& -\langle k \rangle p_k - {kp_k \over \langle k \rangle} \nonumber \\
     &\approx& -(1+k)p_k \quad .      \label{s}
\eea
Here the first term is the contribution of the daughters, while the second  is due to 
the mother. This assumes that the target is not the root.  For simplicity we shall neglect that possibility in the following, which 
makes errors of ${\cal O}(1/N)$.  These are negligible for large $N$.
The last line follows from $\langle k \rangle = 1-1/N \approx 1$, which 
is a good approximation for the same reason.
  
\item $q_k$ is a gain term arising from the possibility that the target acquires new degree $k$. 
Assume that the old degree of the target was $m$, that the degrees of its daughters were $k_1,\ldots , k_m$, 
and that the degree of its mother was $k_0$ --- and that all degrees are uncorrelated. Then 
\bea
   q_k = \sum_m p_m\sum_{k_0\ldots k_m} {k_0p_{k_0} \over \langle k\rangle}\;\prod_{i=1}^mp_{k_i} \;\;
        \delta_{k_0+\ldots k_m-1,k} \;.
\eea
This term is not very transparent. For a more tractable formulation we use 
the generating function methods discussed next.
\end{itemize}

\subsection{Generating Functions}

The generating function for $p_k$ is 
\begin{equation}
   G(x)= \sum_k p_k x^k\quad ,
\end{equation}
and moments of the distribution are given by
\begin{equation}
  \langle k^m \rangle = \left [ \left (x \frac{d}{dx}\right )^m G(x)\right ]_{x=1} \quad .
\label{eq:moments}
\end{equation}
Similarly, the generating function for the gain term is 
\begin{equation}
  Q(x)=\sum_k q_k x^k \quad .
\end{equation}

If a variable has a given generating function, then the generating function for the sum of that variable over $m$ independent realizations is given by the $m^{th}$ power of that generating function~\cite{newman01}. Hence, if the target node has degree $m$, the generating function for the sum of degrees of all its daughters is $[G(x)]^m$.  Using the above definitions and  $G'(1)=\langle k\rangle \approx 1$, we  get 
\be
   Q(x) = \sum_m p_m  G'(x) G^m(x) = G'(x)G(G(x)) \;.
\ee

This, together with Eqs.~(\ref{eq:N}) through~(\ref{s}), leads to 
\be
   \overline{\frac{\Delta G(x)}{\Delta t}} = {1\over N}[G'(x)G(G(x)) - xG'(x)]          
   + {\cal O}(1/N^2) \;\;.        \label{dGdt}
\ee
A more tedious calculation, which requires generating functions for the root of the tree -- arrives at the neglected ${\cal O}(1/N^2)$ terms.  The
 exact result (assuming no correlations)  is
\bea
   \overline{\frac{\Delta G(x)}{\Delta t}} &=& {1\over N}[G'(x)G(G(x)) - xG'(x)] \nonumber \\
          &+& {1\over N^2}[G(G(x))-G(x)] \quad .
\eea
One checks easily that this satisfies the conditions that $G(1)$ is constant and
$G'(1) = 1-1/N$ for all~$t$.

\begin{figure}
\includegraphics*[viewport=30 0 790 600,width=1\columnwidth]{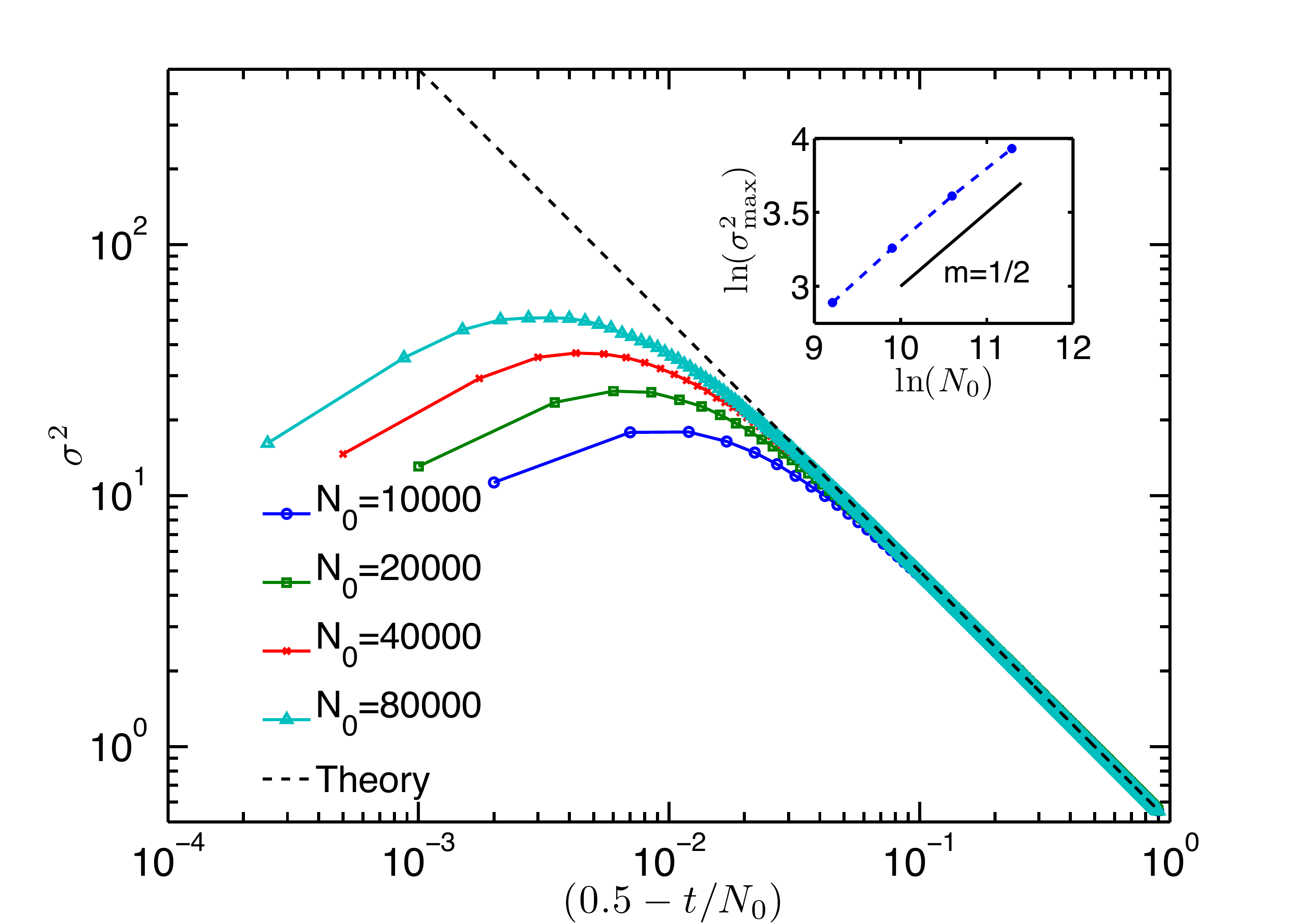}
\caption{(Color online) Comparison between the variance of the degree distribution obtained from 
    Eq.~(\ref{sigma-t}) and simulations for different system sizes, $N_0$. The mean-field theory extends over a larger range  for increasing $N_0$. The inset shows that 
    the maximum variance in RSR observed numerically scales as $N_0^{1/2}$, in agreement with our 
    scaling ansatz Eq.~(\ref{sigma-scaling}).}
\label{fig:2}
\end{figure}

\subsection{Variance of the degree distribution}

Obtaining the time evolution of the variance of the degree distribution requires an expression
for the time evolution of the second derivative of $G$.
From Eq.~(\ref{dGdt})  it follows that
\be
   \overline{\frac{\Delta G''(1)}{\Delta t}} = {2G''(1)\over N} + {\cal O}(1/N^2) \;\;.        \label{dGppdt}
\ee
Making the same steps and approximations as in subsection~\ref{treesize}  gives 
\bea
   {dG''(x) \over dt} &=& {d\langle k^2-k \rangle \over dt} \nonumber \\
&=& { \frac{2\langle k^2-k \rangle}{\langle N\rangle}}
               + {\cal O}(1/\langle N\rangle^2)    \nonumber \\
    &\approx& {2\langle k^2-k \rangle \over N_0-2t}\;.
\eea
Integrating, fixing the integration constant by the condition 
$\langle k^2\rangle_0 = 3/2 + {\cal O}(1/N_0)$, and rewriting the result in terms of the variance
of the degree distribution $\sigma^2$ gives
\be
    \sigma^2 \equiv \langle k^2\rangle - \langle k\rangle^2
              \approx {N_0 \over 2(N_0-2t)} \approx {N_0 \over 2N} \quad .    \label{sigma-t}
\ee

In Fig.~\ref{fig:2} we compare Eq.~(\ref{sigma-t}) for the variance of the degree distribution
with numerical simulations of  RSR
for different initial sizes of critical trees.  We see perfect 
agreement at early times, but increasingly larger disagreement at later times. This
is only in part due to the neglected higher order terms in $1/N$.  Another source of error at late
times is that $N$ exhibits large fluctuations compared to its average. Also,
degree correlations develop. Hence, the mean-field approximation breaks down
for large $t$. But we also see from Fig.~\ref{fig:2} that agreement between theory and numerical results  extends over a broader range for 
increasing system size $N_0$.

\begin{figure}
\includegraphics*[viewport=20 0 778 603,width=8cm,height=6.89cm]{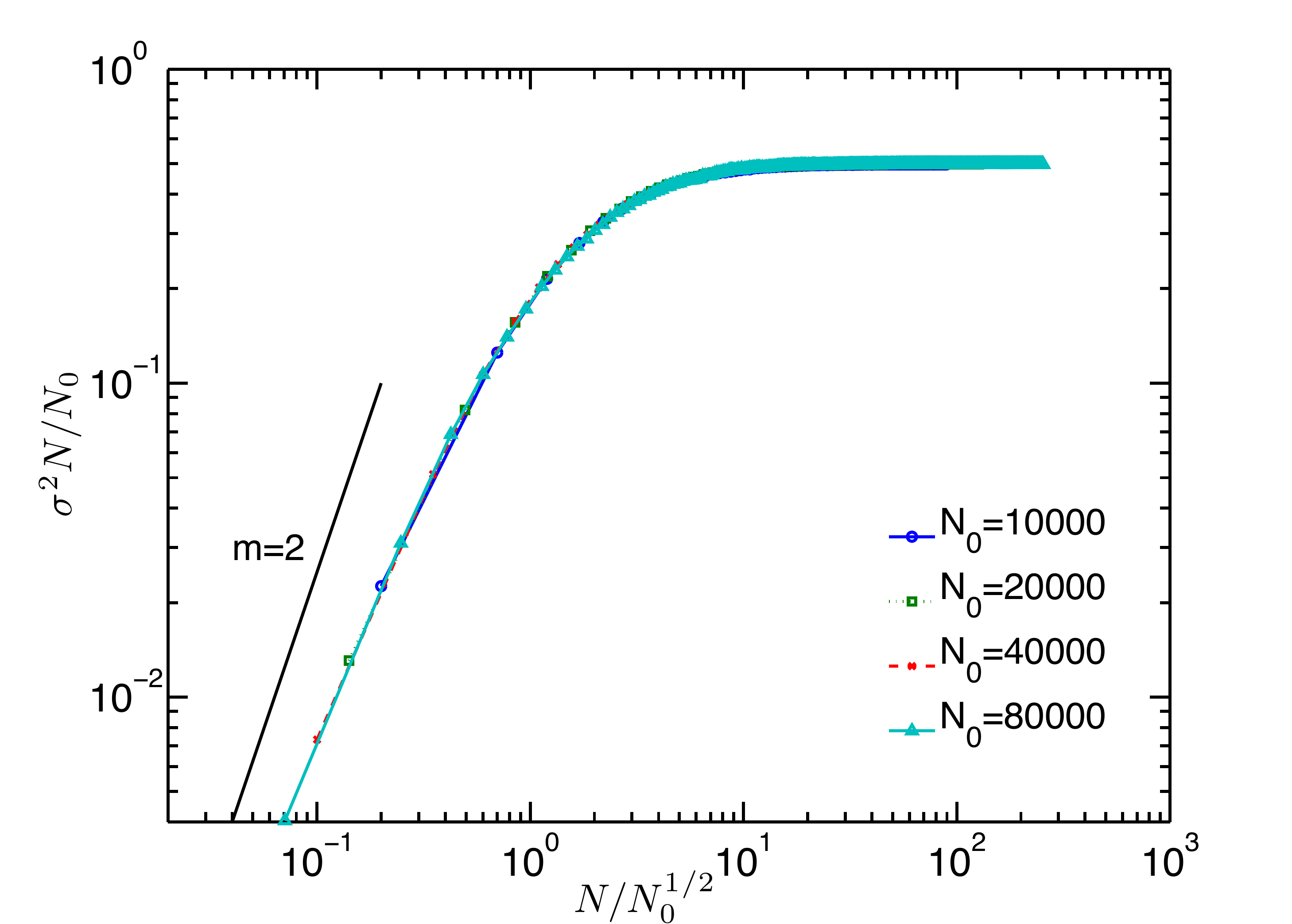}
\caption{(Color online) Scaling of the variance of the degree distribution obtained from RSR. The data are 
   the same as in Fig.~\ref{fig:2}, but the axes are different. They are chosen according to the 
   scaling ansatz Eq.~(\ref{sigma-scaling}), and give excellent data collapse. The straight line has slope
$m=2$.}
\label{fig:3}
\end{figure}

To understand better the behavior at late times (small $N/N_0$), we 
replot the same data using a finite-size scaling (FSS) method in Fig.~\ref{fig:3}. This plot demonstrates that 
the scaling  ansatz
\begin{equation}
    \sigma^2 = \frac{N_0}{N}g\Bigl(\frac{N}{N_0^\nu}\Bigr) \quad ,
                        \label{sigma-scaling}
\end{equation}
with scaling exponent $\nu=1/2$ gives excellent data collapse. We 
derive the result  $\nu=1/2$ in the next subsection.  The scaling function $g(x)$ 
satisfies $g(x) \to 1/2$ for 
$x\to \infty$, in agreement with Eq.~(\ref{sigma-t}).    In addition, the network must,
by definition, end up as a star before it collapses.  Assuming that the star consists of a central hub surrounded by low degree nodes (which is verified numerically), its variance will scale with its size as $\sigma^2 \sim N$. Also, the variance of the degree distribution of the star must be independent of  the initial size $N_0$.    These considerations lead to the conclusion that $g(x) \to x^2$ as $x\to 0$.  Finally, in the scaling ansatz, $g$ and its derivative are continuous functions. 
As a result the maximum variance occurs when $N\sim N_0^{1/2}$ so that
the maximum value of $\sigma^2 \sim N_0^{1/2}$, in agreement with the inset of Fig.~\ref{fig:2}.
Scaling laws like Eq.~(\ref{sigma-scaling}) 
in terms of homogeneous functions are well known from critical phenomena~\cite{amit,Zinn},
where they describe finite-size scaling  with several control 
parameters such as temperature and magnetic field.

\subsection{Fluctuations of the system size and the relaxation time}
\label{random_walk_sec}
In this subsection we  derive the result $\nu=1/2$ by considering fluctuations around the
average value of $\Delta N/\Delta t$, and the resulting fluctuations both of $N_t$
and of the relaxation time $T$.  (Recall that the  latter is defined as the time when the tree is 
first reduced to a single node.) Here we explicitly label the fluctuating number of nodes with its
time dependence $N_t$.

Generalizing Eq.~(\ref{eq:N}) and neglecting the ${\cal O}(1/N)$ term, we make the ansatz
\be
   {\Delta N_t\over \Delta t} = -2+ \epsilon_t \quad .
\label{eq:Nfluc}
\ee
Here $\epsilon$ is a random variable with zero mean and with variance equal to the 
variance of the degree distribution $\sigma^2_t$, which, on average, increases with time $t$. 
Assuming no degree correlations, the random variables $\epsilon_t$ at different 
times are also uncorrelated, and
\be
   \langle \epsilon_t \epsilon_{t'}\rangle = \delta_{t,t'} \sigma^2_t \;.
\ee
Thus the fluctuations of $N_t$ are given by 
\be
   \delta N_t \equiv N_t - \langle N_t\rangle = \sum_{t'=0}^{t-1}\epsilon_{t'}.
\ee
Since $\sigma_t$ is finite for all $t$, the central limit theorem implies that 
$\delta N_t$ is Gaussian for large $t$ with variance 
\bea
   {\rm Var}[\delta N_t] &=& \sum_{t'=0}^{t-1}\sigma^2_{t'} 
         \approx \sum_{t'=0}^{t-1}{N_0 \over 2(N_0-2t')} \nonumber \\
       &\approx& {N_0\over 4} \ln {N_0\over \langle N_t\rangle} \;.
\eea 

This estimate has to break down when typical fluctuations of $N_t$ are as big as 
its average,  or when ${\rm Var}[\delta N_t] \approx \langle N_t\rangle^2$. We 
claim that this happens at a time when $\langle N_t\rangle\sim N_0^{1/2}$, explaining the fact that
$\nu=1/2$. Indeed, when $\langle N_t\rangle \sim N_0^\nu$ with some positive exponent $\nu$,
then ${\rm Var}[\delta N_t] \sim N_0 \ln N_0 > N_0$ for large $N_0$,
implying that it is larger than $\langle N_t\rangle^2$ for any $\nu < 1/2$. On the other hand,
${\rm Var}[\delta N_t]$ increases less quickly than $\langle N_t\rangle^2$ for any $\nu >1/2$,
showing that the initial scaling regime breaks down when $\langle N_t\rangle\sim N_0^\nu$ 
with $\nu=1/2$.

\begin{figure}
\includegraphics*[viewport=30 0 785 570,width=8cm,height=6.5cm]{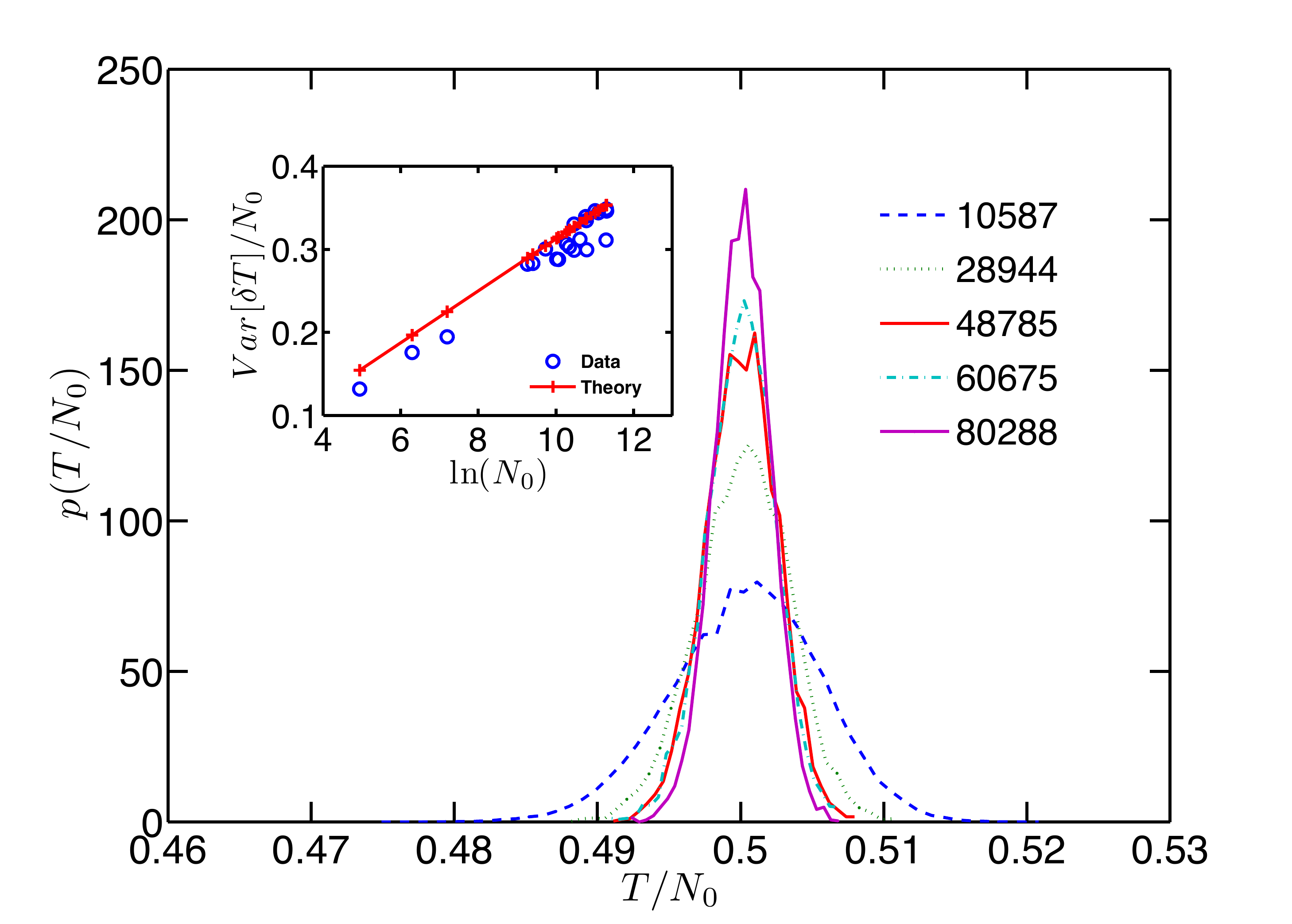}
\caption{(Color online) Distributions of relaxation times for various values of $N_0$. 
   The inset compares the variance of these distributions to 
   Eq.~(\ref{eq:deltaT}) finding good agreement. }
\label{fig:4}
\end{figure}

Fluctuations of the relaxation time $T$ are obtained by demanding that $N_T=1$,
which gives 
\be
    2T - \sum_{t'=0}^{T-1}\epsilon_{t'} = N_0 \;.
\ee
Hence, for large $N_0$, $T$ is distributed as an inverse Gaussian variate which is 
well approximated in the large $N_0$ limit by an ordinary Gaussian. Strictly, its variance cannot be 
calculated exactly, since the summation extends beyond the limit of applicability
of our theory. To take this into account, we first convert the summation over $t'$ to an integral
over $N$ and truncate the integral at $N_0^{1/2}$, where the mean-field theory breaks down. 
Integration gives
\be
   {\rm Var}[\delta T] ={1\over 32} N_0 \ln N_0\;   \label{eq:deltaT}
\ee
plus lower order terms.
This is compared with the simulation results shown in the inset of Fig.~\ref{fig:4}, finding
good agreement.  

\subsection{Scaling of maximum degree}

A simple way to track the formation of hubs under RSR is to measure the
maximum degree in the network $k_{\rm max}$.  A naive scaling assumption is that when a few 
large hubs together with many low degree nodes 
dominate,  $\sigma^2 \sim k_{\rm max}^2/N$.  Using Eq.~(\ref{sigma-scaling}) gives
\be
    k_{\rm max} \sim N_0^{1/2} f\left(\frac{N}{N_0^{1/2}}\right) \quad \; .\label{eq:kmax}
\ee
\begin{figure}
\includegraphics*[viewport=30 0 785 586,width=8cm,height=6.75cm]{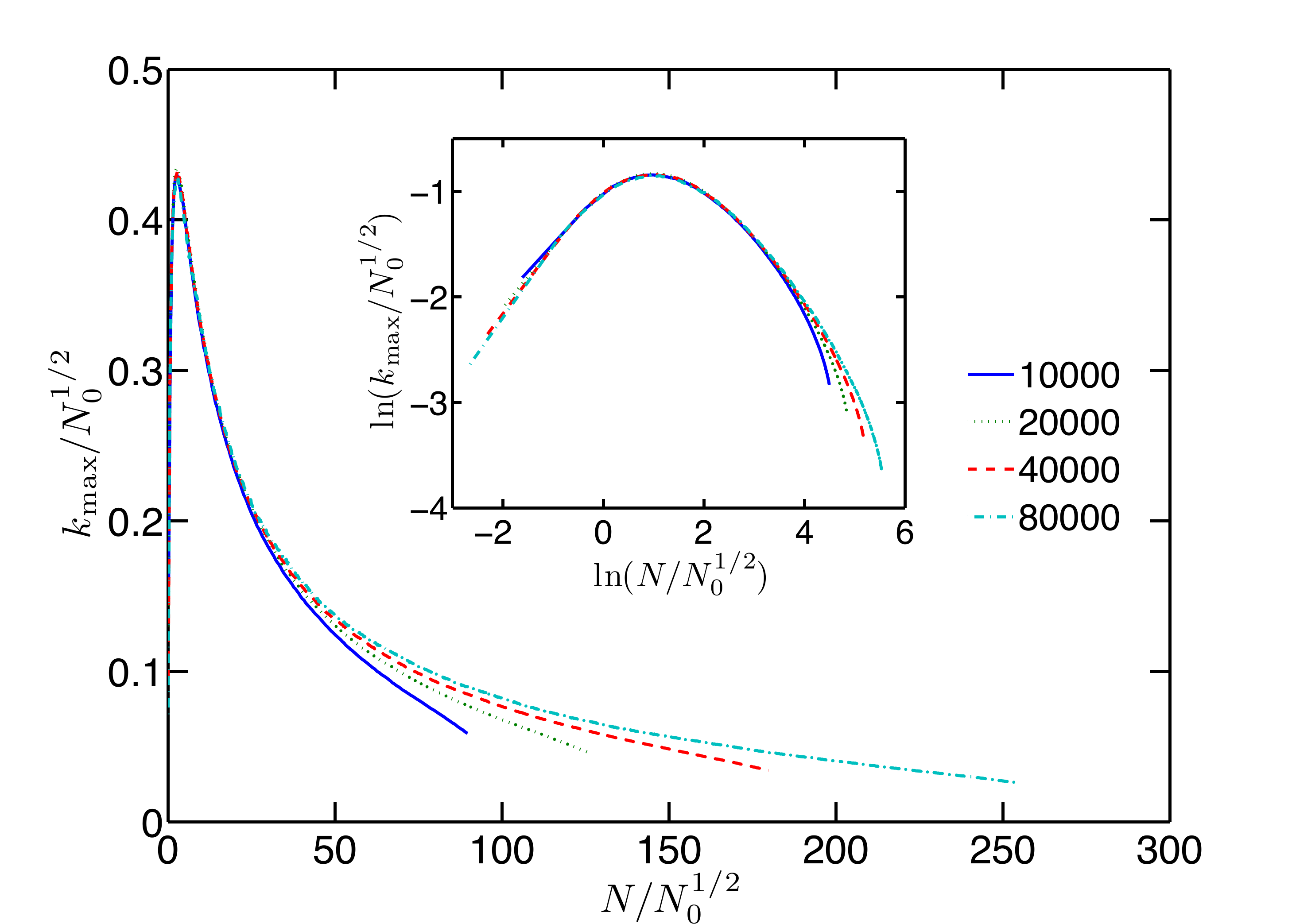}
\caption{(Color online) FSS analysis of $k_{\rm max}$ using Eq.~(\ref{eq:kmax}).  There is  perfect data collapse in the region $N\sim N_0^{1/2}$.  }
\label{fig:kmax}
\end{figure}
Figure~\ref{fig:kmax} compares this equation to results from numerical simulations.
While there are clear (and expected) deviations  for $N/N_0^{1/2}\to\infty$,
the collapse in the intermediate region $N\sim N_0^{1/2}$, where $\sigma^2$ achieves
its maximum, is perfect.  As before, assuming that the tree evolves to a star with a hub
at its center suggests that $f(x)\sim x$ as $x\to 0$.  
However in Fig.~\ref{fig:kmax} we do not observe this behavior as the fitting region is small
and there is still some curvature in the scaling function.  As for $\sigma^2$, our  theory
predicts that the largest value of $k_{\rm max}$ observed under RSR scales as $N_0^{1/2}$ and agrees with the data seen in the inset of Fig.~\ref{fig:kmax}.

\begin{figure}
\centering
\includegraphics*[viewport=60 0 770 600,width=8cm]{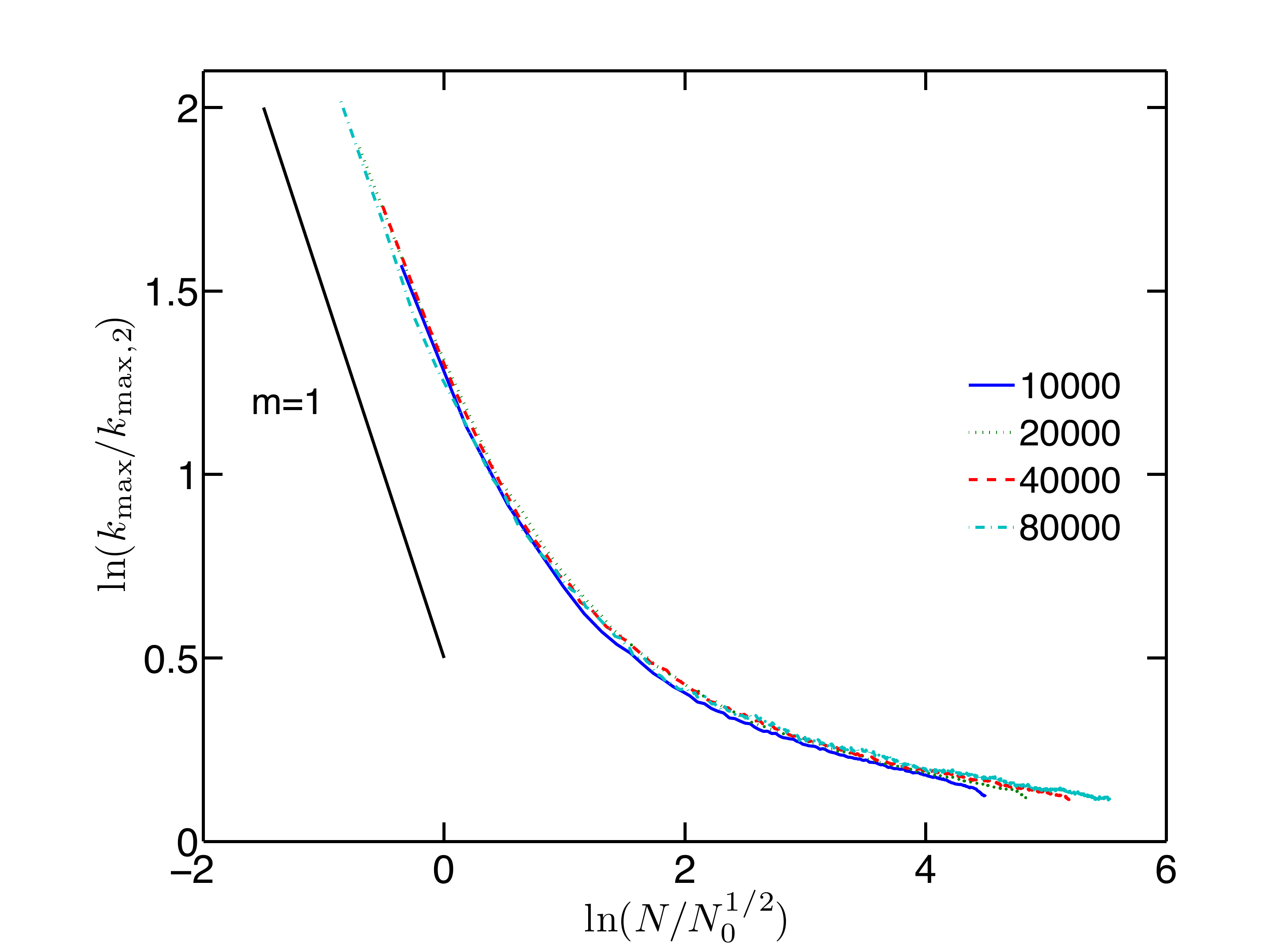}
\caption{(Color online) FSS analysis of $k_{\rm max}/k_{\rm max,2}$.  This ratio increases as one large hub 
    separates from the rest of the degree distribution. The data show good agreement with the scaling 
   ansatz Eq.~(\ref{eq:kmaxratio}). The line with slope $-1$ indicates the theoretical 
    prediction as the network approaches a star. 
}
\label{fig:ratio}
\end{figure}

\subsection{Ratio of the largest degree to the second largest degree}
The ratio of $k_{\rm max}$ to the second largest degree $k_{{\rm max},2}$  (provided that  $k_{{\rm max},2}>0$) is shown in 
Fig.~\ref{fig:ratio}. It agrees with an FSS analysis using the same 
exponent $\nu=1/2$,
\be
    {k_{\rm max}\over k_{{\rm max},2}} = h\left(\frac{N}{N_0^{1/2}}\right) \quad  .
    \label{eq:kmaxratio}
\ee
Once again the extreme limits of the scaling function $h$ can be determined.  For the initial network the
largest and second largest degree are equal, so $h(x\to \infty)\to 1$.  For a pure star of size $N$,
$k_{\rm max}/k_{{\rm max},2} = N$. As shown in Section~\ref{last_size_sec}, stars first appear when
$N\sim N_0^{\nu_{\rm star}}$ with $\nu_{\rm star}\approx 1/4$.  In that case $k_{\rm max}/k_{{\rm max},2}\sim N_0^{1/4}$. Hence $h(N_0^{-1/4})\sim N_0^{1/4}$, or
$h(x\to 0)\sim 1/x$. Figure~\ref{fig:ratio} shows that $h$ is increasing in this limit, although the asymptotic
regime is not yet reached for the system sizes studied.

\subsection{Degree distribution}
          \label{degdist}

\begin{figure}
\includegraphics*[width=\columnwidth,height=7cm]{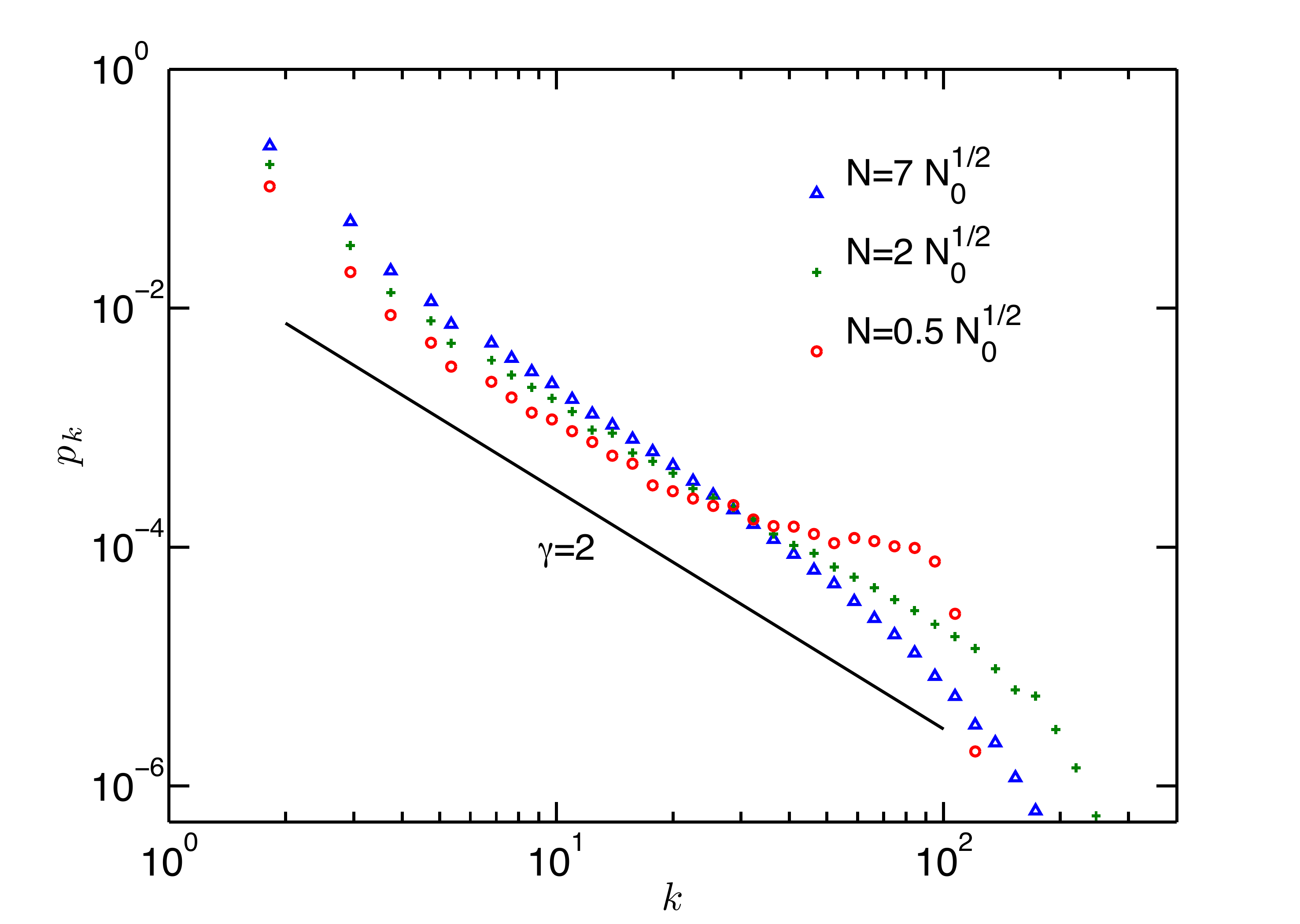}
\caption{(Color online) Log-log plot of the degree distribution $p_k$ for trees with $N_0=8\times 10^4$ 
   at three values of $N$: $N = 7N_0^{1/2}$, $N = 2N_0^{1/2}$ and $N = 0.5 N_0^{1/2}$.  These distributions
   are obtained by averaging over different initial networks and different realizations of  RSR. The distribution widens and then becomes more narrow on decreasing $N$ as
   hubs separate from the rest of the nodes during the transition.
   The data are consistent with our theoretical prediction that at the critical point 
   $p_k\sim k^{-\gamma}$  with $\gamma=2 $.}
\label{fig:degdist}
\end{figure}
Degree distributions for large initial trees at three  points in the evolution 
are shown in Fig.~\ref{fig:degdist}. 
Critical trees start with a narrow degree distribution, which becomes broader and 
broader under RSR. The degree distribution gradually transforms into a power law distribution as $N$ 
approaches $\sim N_0^{1/2}$.   For a power law degree distribution $p(k)\sim k^{-\gamma}$, the
variance obeys
\begin{equation}
\sigma^2 \sim \int^{k_{\rm max}} k^{2-\gamma} dk \sim k_{\rm max}^{3-\gamma}\quad .
\end{equation}
From the  scaling result at the transition, $\sigma^2 \sim k_{\rm max} \sim N_0^{1/2}$, we get
$\gamma =2$, consistent with the data shown.
 
With the formation of a giant hub at the transition, 
a bump appears at large $k$ in $p_k$. This is clearly visible for $N=0.5N_0^{1/2}$ in
Fig.~\ref{fig:degdist}. Note that the distributions shown in this figure are obtained by 
averaging over many initial networks and many realizations of RSR. In the degree 
distribution of a single network a gap emerges between the largest hub and 
the rest of the nodes, for $N\sim N_0^{1/2}$ as demonstrated in  Fig.~\ref{fig:ratio}.

\begin{figure}[!h]
\includegraphics*[width=\columnwidth,height=7cm]{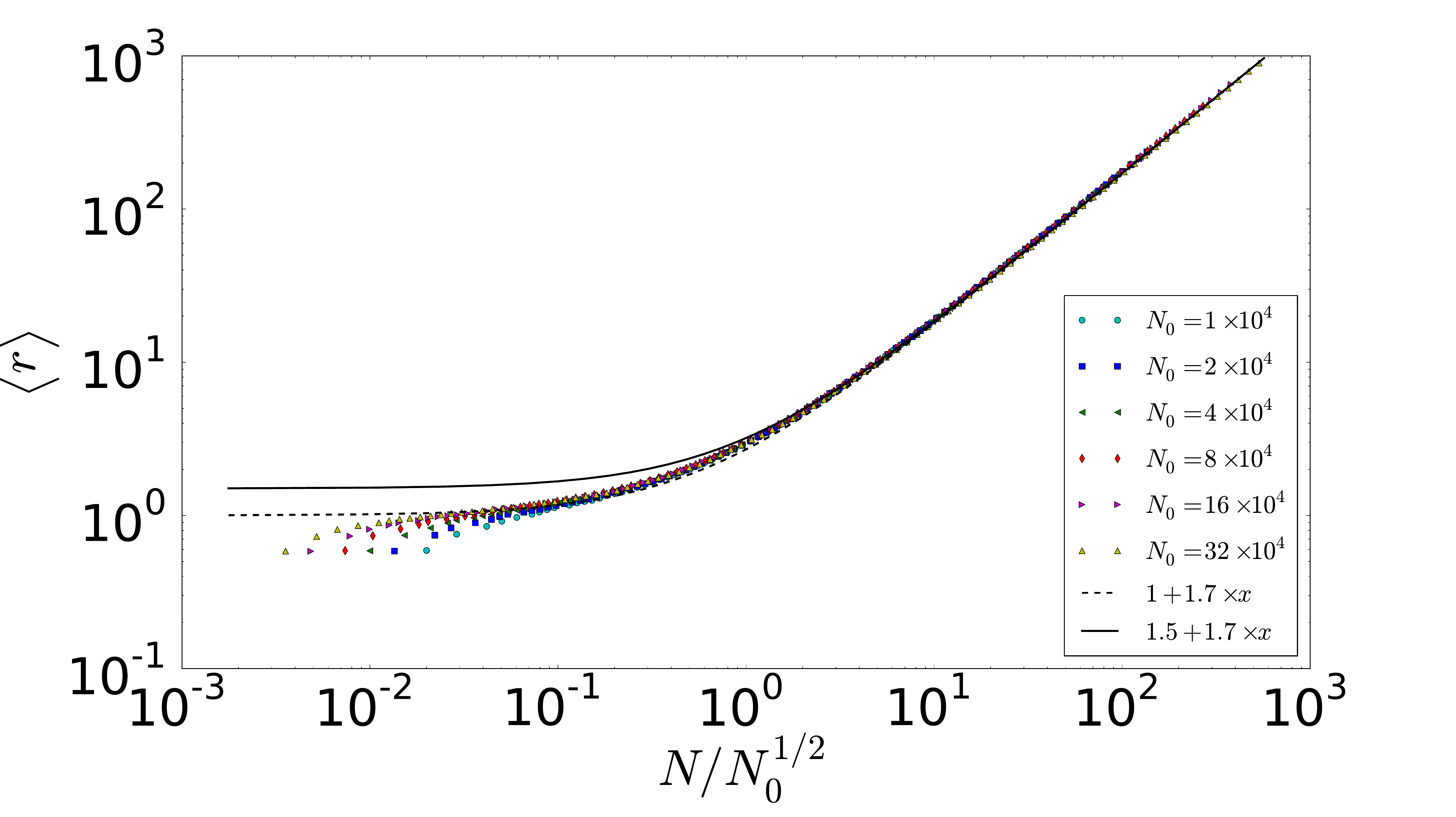}

\vskip -.5cm
\caption{(Color online) Mean radius $r$ of trees as a function of $N/N_0^{1/2}$. Agreement with Eq.~\ref{eq:bounds} is excellent with $\alpha=1.7$ as indicated. Fluctuations cannot be ignored for small $N/N_0^{1/2}$ when mean-field theory breaks down and the bounds are no longer valid.}
\label{fig:mean_radius}
\end{figure}
\begin{figure}[!h]
\includegraphics*[width=\columnwidth,height=7cm]{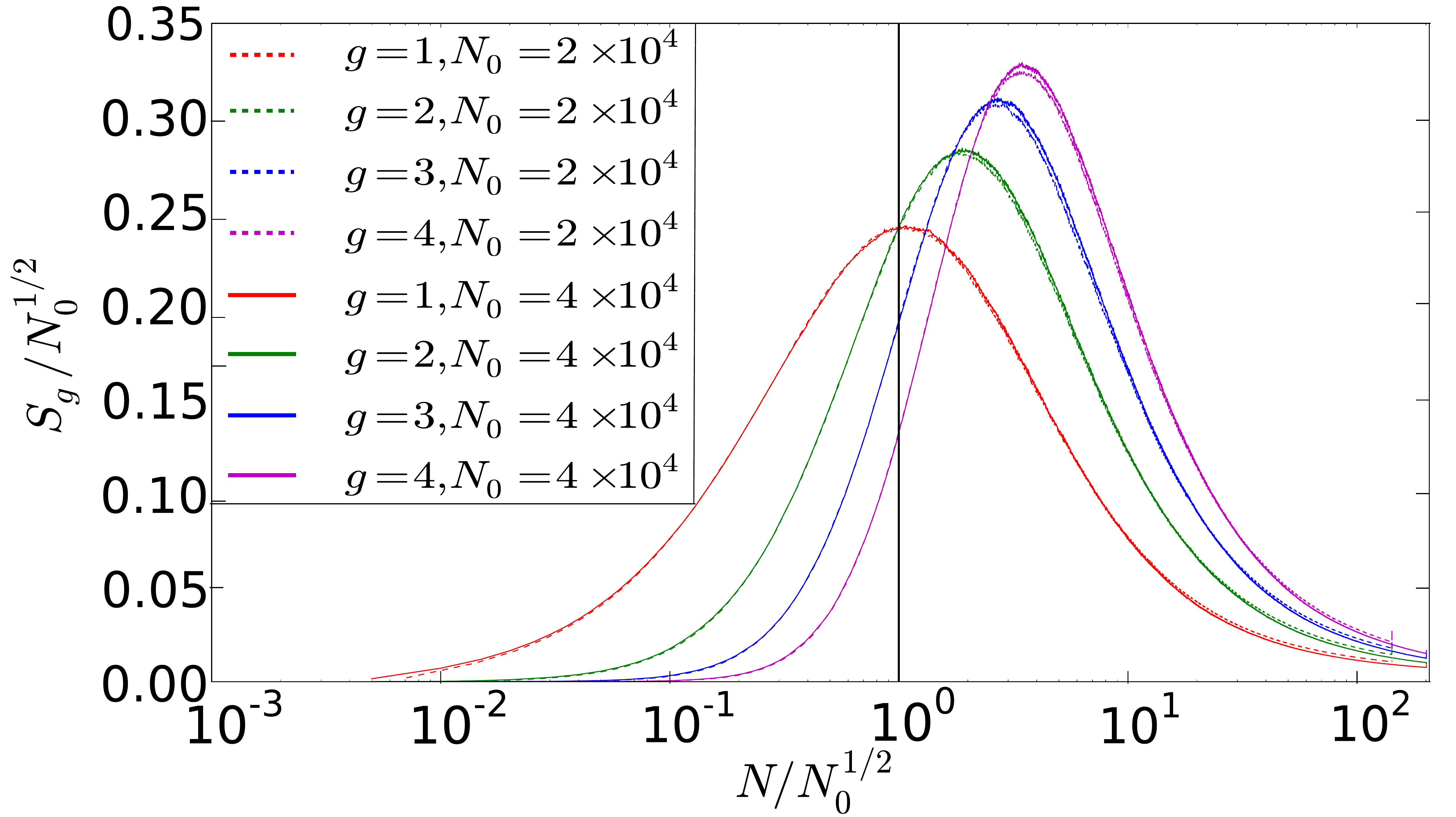}
\caption{(Color online) The evolution of the number of nodes in the first four shells as a function of $N/N_0^{1/2}$ for two different system sizes.  Note that $S_1$ crosses $S_2$ to become the largest shell at $N=N_0^{1/2}$.  The other shells vanish increasingly faster as $N$ decreases further.}
\label{fig:shell_sizes}
\end{figure}
\subsection{Mean-field theory for average  radius of trees}

The sum of the distances of nodes from the root in a tree of size $N$ can be written as
\be
R=\sum_{x=1}^{N-1} g_x \quad ,
\ee
where $g_x$ is the distance of node $x$ from the root.
It is simplest to consider that (except for the root)  the mother of a target node absorbs her (target) daughter plus all of that daughter's daughters. Consider node $x$ at distance $g_x>1$.  If the root is the target in the next RSR step, $g_x$ is reduced by $1$.  If an ancestor of $x's$ mother is hit, which is not the root, then $g_x$ is reduced by $2$.  If either $x$ or her mother is the target, then $x$ disappears, contributing zero to $R$. Hence the position of $x$ evolves in the continuous time approximation on average as
\be
N{\partial g_x\over \partial t}= -1 -2 (g_x -2) -2g_x = -4g_x +3 \quad ,
\ee
for $x>1$. For $x=1$
\be
N{\partial g_x\over \partial t}= -2 \quad .
\ee
We can write the evolution in terms of the average number of nodes instead of time.  As before, in mean-field we ignore fluctuations in $N$ about its average $\langle N\rangle$, in $R$ about its average $\langle R\rangle$, and in the number of
nodes at distance $1$ in the tree $S_1$ about its average $\langle S_1\rangle$.  This gives, after dropping
all angular brackets,
\be
{dR \over dN} = {2R\over N} - {3\over 2}(1- {1\over N}) + {S_1\over 2N} \quad .
\label{eq:RoverN}
\ee
Defining the average radius $r=R/N$ with  initial value $r_0=\alpha N_0^{1/2}$ for large $N_0$, the constant
 $\alpha \sim {\cal O}(1)$ depends on the precise rule for constructing  critical trees. Equation~(\ref{eq:RoverN}) can be solved to get
\be
r(N) = {3\over 2}\Bigl(1 -{N\over N_0}\Bigr)  +\alpha{N\over N_0^{1/2}}  - {N\over 2}\int_N^{N_0} dy\Bigl({S_1\over y^3}\Bigr) \; .
\ee
Bounds on $r(N)$ can be placed based on the fact that $1\leq S_1 <N$ to get
\bea
1 +  \alpha\Bigl({N\over N_0^{1/2}}\Bigr)  - {N\over 2N_0}  <r   \leq \nonumber \\ 
  {3\over 2}  \Bigl(1-{N\over N_0}\Bigr) +  \alpha\Bigl({N\over N_0^{1/2}}\Bigr)  -{1\over 4N} + {N\over 4N_0^2}
\label{eq:bounds}
\quad .
\eea
These bounds are tested against numerical data in Fig.~\ref{fig:mean_radius} showing excellent agreement,
up until the regime where $N$ becomes small compared to $N_0^{1/2}$.  At that point
 mean-field theory breaks down.
As the trees start to exit the mean-field regime, their average radius becomes order unity even for $N\sim N_0^{1/2}\to \infty$.   Figure~\ref{fig:shell_sizes} shows the evolution of the average number of nodes at distances
1, 2, 3, and 4 from the root, ($S_1,\; S_2,\; S_3,\; {\rm and}\; S_4$, respectively).  At $N=N_0^{1/2}$, $S_1$ becomes the largest shell, and $S_2$ seems to be exactly equal to $S_1$ at that point.
All other shells vanish compared to $S_1$ for smaller $N$.  This is the origin of the finite radius of renormalized trees near the end of the mean-field regime.

\subsection{Distribution of last sizes and the star regime}
\label{last_size_sec}
Before the network reaches the trivial
fixed point at $N=1$ it must first turn into a star.  The star eventually collapses into a single node
when the central node is hit as the target.

\begin{figure}
\includegraphics*[viewport=20 0 780 576,width=8cm,height=6.75cm]{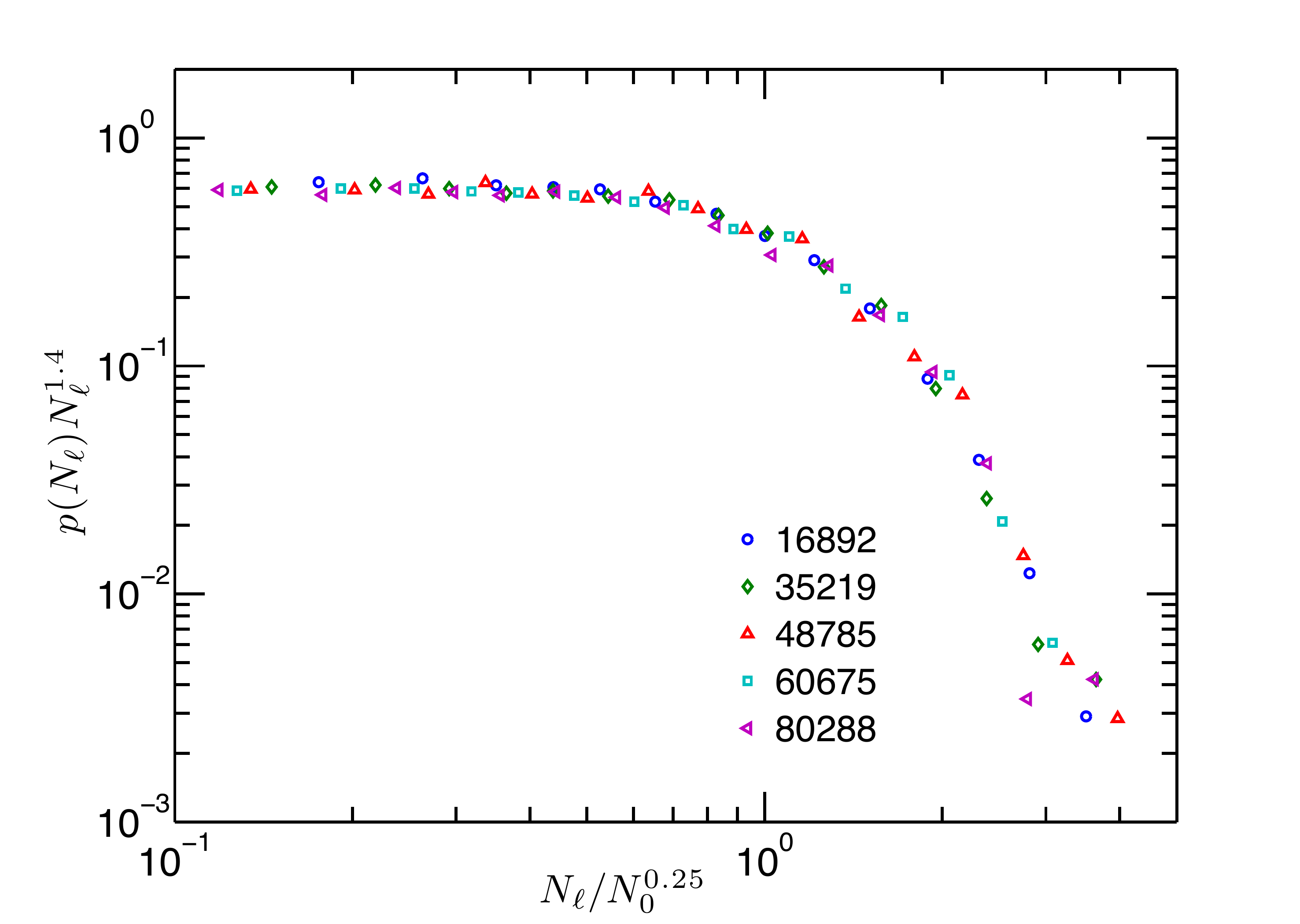}
\caption{(Color online) FSS analysis for the  distribution of last sizes  based on Eqs.~(\ref{eq:pnl})-(\ref{eq:pnl2}) with $\tau=1.4 \pm 0.1$ and $D=0.25 \pm 0.07$. In view of the comment after Eq.~(\ref{eq:star}), $p(N_{\ell})$ is replaced with $p(N_{\ell})/2$ for $N_{\ell}=2$.}
\label{fig:Ctree_Nlast2}
\end{figure}

We define the quantity $N_{\ell}$ to be the size of the network one step before it dies. 
Figure~\ref{fig:Ctree_Nlast2} shows an FSS plot for the probability distribution of $N_{\ell}$.  More 
precisely, it shows $N_{\ell}^{1.4}\;p(N_{\ell})$ against $N_{\ell}/N_0^{1/4}$. The data 
collapse seen suggests a scaling form
\begin{equation}
   p(N_{\ell})\sim \frac{1}{N_{\ell}^{\tau}} \Phi(N_{\ell}/N_0^{D}),
\label{eq:pnl}
\end{equation}
with $\tau=1.4 \pm 0.1$, $D=0.25 \pm 0.05$. The scaling function $\Phi(x)$ seems to approach a constant 
for $x\to 0$, suggesting that $p(N_{\ell})$ tends to a power law, 
$p(N_{\ell})\sim N_{\ell}^{-\tau}$, for $N_{\ell} \ll N_0^{1/4}$.

From the distribution of $N_{\ell}$ we can determine the distribution of sizes 
when the tree first turns into a star. Let us call $s$ the size when the renormalized tree  first 
reaches a star configuration, and $p_s(s)$ its distribution. 
In each subsequent time step the star can either shrink by
exactly one node (probability $1-1/N$), or it can be reduced immediately to a single 
node (probability $1/N$). Starting with a star of size $s$, the conditional probability to end up at 
final size $N_{\ell}$ is 
\be
p(N_\ell|s) = \left\{
\begin{array}{l c c}
  {\displaystyle  { 1\over s}} & ,  &N_\ell=s \\
  \\
  \displaystyle \prod_{t=1}^{s-N_\ell} \frac{s-t}{s-t+1}\frac{1}{N_\ell} =\frac{1}{s}  \quad & , &  2<N_\ell<s \\
  \\
  {\displaystyle  { 2\over s}} & , & N_\ell=2  
\end{array}
\right.
\label{eq:star}
\ee
where the last line comes from the degeneracy of a star with two nodes and is required for proper normalization.  Assuming  that $p_s(s)$ has a scaling
form with possibly new exponents and a new scaling function $\phi$,
\be
   p_s(s)\sim \frac{1}{s^{\alpha}} \phi(s/N_0^\beta) \;,                    \label{eq:ps}    
\ee
we obtain 
\bea
   p(N_{\ell}) &=& \sum_{s\geq N_{\ell}} p(N_{\ell}|s)p_s(s) \nonumber \\
               &\approx& \int_{N_{\ell}}^\infty ds {\phi(s/N_0^\beta)\over s^{1+\alpha}} \nonumber \\ 
               &=& {1\over N_\ell^\alpha} \Psi(N_{\ell}/N_0^\beta)
\label{eq:pnl2}
\eea
with $\Psi(x)=x^{\alpha}\int_x^\infty dx'\;\phi(x')/x'^{1+\alpha}$. This agrees 
with Eq.~(\ref{eq:pnl}), if we identify $\alpha = \tau$, $\beta = D$, and 
$\Psi(x)=\Phi(x)$. Thus the distributions of $s$ and of $N_\ell$ have the 
same exponents, if they obey FSS, which we verified numerically.

\section{Conclusion}

To study invariant properties of graphs under coarse graining, we have
 introduced the random sequential renormalization (RSR) method, where 
in each step only a part of the network within a fixed distance $b$ from a randomly chosen node 
collapses into one node. RSR is easy to implement and eliminates the problem of finding an
optimum tiling of the  network.  In addition, the small effect of each decimation  gives a much 
more detailed statistical picture of  the renormalization flow. We applied the RSR 
with $b=1$ to critical trees and  derived results analytically, finding good agreement
with  numerical simulations.

Under renormalization a critical regime appears when the size of the tree $N \sim N_0^{\nu}$
with $\nu=1/2$.   The behavior of the tree before this regime is reached is described using 
a mean-field theory based on generating functions. There is a constant $c\simeq 1$ such that the 
degree distribution of the network is scale free, $p_k\sim k^{-\gamma}$ with $\gamma=2$, in the 
limit $N_0\to \infty$ and $N/N_0^{1/2}=c$. Both the variance of the degree distribution $\sigma^2$ 
and the maximum degree in the network $k_{\rm max}$ diverge as $N_0^{1/2}$ in this limit. Both of 
these quantities are described by crossover functions exhibiting finite-size scaling that 
connect the mean-field regime to a regime for $N_0^{1/4}\lesssim N \lesssim N_0^{1/2}$ when 
hubs start to emerge.  Results from numerical simulations agree with a scaling theory we develop 
to describe this fixed point.  Trees are short and fat near this point with an average depth ${\cal O}(1)$.
As RSR proceeds further, star configurations start to appear for 
$N\sim N_0^{\nu_{\rm star}}$ with $\nu_{\rm star}\approx 1/4$.  The distribution of star sizes seems to obey FSS,  characterized by its own critical exponents,
which we were not able to derive analytically.

We began this investigation to study in a more controlled way claims made in 
the literature about real-space renormalization of complex networks \cite{Song,Radi1,Radi2}. 
In the most detailed previous study~\cite{Radi1,Radi2} many of the findings are similar to ours, 
with the caveat that unlike previous works, the results presented here are for critical  
trees rather than for general networks.  The most striking and robust agreement is the 
emergence of hubs under renormalization --  which leads to a final star regime. Associated 
with the emergence of hubs is a fixed point that gives rise to a power law degree distribution.

An alternative way to describe RSR is the following: Instead of removing nodes in each coarse 
graining step and replacing them by a 
new ``super''-node, we keep them and join them into a cluster. At each subsequent RSR step, 
entire clusters are joined into new ``superclusters."  This process, where clusters grow by attaching to all the neighbors is an aggregation process~\cite{Son-2010} is called ``agglomerative percolation'' (AP)  in Ref.~\cite{claire}.
The original network has only clusters of 
size one, but larger and larger clusters appear as the RG flow goes on. At the critical point, an
infinite cluster (in the limit $N_0\to\infty$) appears. In this interpretation, the critical 
behavior seen in this paper (and in Refs.~\cite{Radi1,Radi2}) is just a novel type of percolation.

If the original network is a simple chain, the probability 
 distribution to find any sequence of masses for any
 $b$, initial size $N_0$, and time $t$ have been derived exactly.
In this case, AP exhibits critical exponents different from ordinary percolation. These exponents depend
on $b$~\cite{Son-2010}.  In two dimensions on a square lattice, AP is in a different universality class than ordinary percolation~\cite{claire}.

In future work~\cite{nextpaper} we plan to study RSR on networks that are more complex than trees. 
For Erd\"os-Renyi graphs we have found a fixed point at finite ratio $N/N_0$ associated with the emergence 
of hubs, which in the case of critical trees and of simple chains is driven to zero. 
This difference between trees and 
Erd\"os-Renyi graphs is intuitively most easily understood in the percolation picture discussed 
above. Trees having topological dimension one, any percolation transition on them can only happen 
when the probabilities for establishing bonds goes to one.

It remains to be seen whether RSR  (or equivalently AP)
 can be used as a generic tool to uncover universality classes 
in large networks (in the usual renormalization group sense) by eliminating irrelevant degrees 
of freedom.  On a more speculative note, our results point to another way to create scale free 
networks that is not based on an explicit generative mechanism for power law behavior at the 
microscopic scale, but result from hubs being aggregates of many microscopic  nodes. That would
suggest the view that networks are emergent collections of 
smaller networks made up of even smaller ones down to the lowest scales. 

Acknowledgements: We thank Claire Christensen for very helpful discussions, in particular for 
pointing out the connection with percolation processes.

\bibliographystyle{apsrev4-1}
\bibliography{bib1}

\end{document}